\documentclass[preprint2,numberedappendix]{emulateapj-rtx4}

\usepackage{graphicx,natbib,bm,color}

\graphicspath{{./fig/}{./png/}}

\newcommand{\kk}{\bm{k}}
\newcommand{\xx}{\bm{x}}
\newcommand{\G}{\,{\rm G}}
\newcommand{\kG}{\,{\rm kG}}
\newcommand{\cm}{\,{\rm cm}}
\newcommand{\km}{\,{\rm km}}
\newcommand{\Mm}{\,{\rm Mm}}
\newcommand{\Mx}{\,{\rm Mx}}
\newcommand{\erg}{\,{\rm erg}}
\def\half{{\textstyle{1\over2}}}

\newcommand{\Eq}[1]{Equation~(\ref{#1})}

\newcommand{\Fig}[1]{Figure~\ref{#1}}
\newcommand{\FFig}[1]{Figure~\ref{#1}}

\newcommand{\Tab}[1]{Table~\ref{#1}}

\newcommand{\ybook}[3]{ #1, {#2} (#3)}
\newcommand{\yapj}[3]{ #1, {ApJ,} {#2}, #3}
\newcommand{\sapj}[2]{ #1, {ApJ}, submitted, arXiv:#2}
\newcommand{\yapjl}[3]{ #1, {ApJL,} {#2}, #3}
\newcommand{\yana}[3]{ #1, {A\&A,} {#2}, #3}
\newcommand{\yjfm}[3]{ #1, {J.\ Fluid Mech.,} {#2}, #3}
\newcommand{\ypre}[3]{ #1, {Phys.\ Rev.\ E,} {#2}, #3}
\newcommand{\yprl}[3]{ #1, {Phys.\ Rev.\ Lett.,} {#2}, #3}
\newcommand{\ysph}[3]{ #1, {Solar Phys.,} {#2}, #3}
\newcommand{\yasr}[3]{ #1, {Adv. Space Res.,} {#2}, #3}
\newcommand{\yjaa}[3]{ #1, {J. Astrophys. Astr.,} {#2}, #3}
\newcommand{\ypasj}[3]{ #1, {Publ. Astron. Soc. Japan,} {#2}, #3}
\newcommand{\ymnras}[3]{ #1, {Mon. Not. R. Astron. Soc.,} {#2}, #3}
\newcommand{\yjour}[4]{ #1, {#2}, {#3}, #4}

\begin{document}

\title{Spectral magnetic helicity of solar active regions between 2006 and 2017}

\author{
Sanjay Gosain$^1$\thanks{E-mail:sgosaiw@gmail.com} \&
Axel Brandenburg$^{2,3,4,5}$
}

\affil{
$^1$National Solar Observatory, 3665 Discovery Drive, Boulder, CO 80303, USA\\
$^2$Nordita, KTH Royal Institute of Technology and Stockholm University, Roslagstullsbacken 23, SE-10691 Stockholm, Sweden\\
$^3$Department of Astronomy, AlbaNova University Center, Stockholm University, SE-10691 Stockholm, Sweden\\
$^4$JILA and Laboratory for Atmospheric and Space Physics, University of Colorado, Boulder, CO 80303, USA\\
$^5$McWilliams Center for Cosmology \& Department of Physics, Carnegie Mellon University, Pittsburgh, PA 15213, USA
}

\date{~$\! $Revision: 1.75 $ \!$}

\begin{abstract}
We compute magnetic helicity and energy spectra from about 2485 patches
of about $100$ megameters (Mm) side length on the solar surface using data
from {\em Hinode} during 2006--2017.
An extensive database is assembled where we list magnetic energy and
helicity, large- and small-scale magnetic helicity, mean current helicity
density, fractional magnetic helicity, and correlation length along with
the {\em Hinode} map identification number (MapID), as well as Carrington
latitude and longitude for each MapID.
While there are departures from the hemispheric sign rule for magnetic and
current helicities, the weak trend reported here is in agreement with the previous results.
This is argued to be a physical effect associated with the dominance
of individual active regions that contribute more strongly in the better resolved {\em Hinode} maps.
In comparison with earlier work, the typical correlation length is
found to be $6$--$8\Mm$, while the length
scale relating magnetic and current helicity to each other is found to
be around $1.4\Mm$.
\end{abstract}

\keywords{
Sun: magnetic fields --- dynamo --- magnetohydrodynamics --- turbulence
}

\section{Introduction}

The Sun's global magnetic field is produced by a large-scale dynamo
where the overall rotation and vertical density stratification are
believed to play important roles in driving what \cite{Par55} called
cyclonic convection.
This means that the flow has a swirl, which can be quantified by its
kinetic helicity.
Although the details of the solar dynamo are still being debated, there
is no doubt that also the Sun's magnetic field possesses helicity.
This was first found by \cite{See90}, who determined the swirl of electric
current lines, i.e., the current helicity, as the product of the vertical
components of magnetic field and current density.
Its value was found to be predominantly negative in the northern
hemisphere and positive in the southern.

Subsequent work by \cite{PCM95} confirmed the overall hemispheric
dependence, but also showed significant scatter.
The work of \cite{BZ98} using the Huairou Solar Observing Station of
the Beijing Astronomical Observatory also showed scatter, but it was
less than what was found by \cite{PCM95}.

The study of solar magnetic helicity received wide-spread attention
with the Chapman Conference in Boulder/Colorado during July 28--31,
1998 \citep{BCP99}.
Nowadays, the most commonly employed methods for quantifying
magnetic swirl or twist in the Sun include the determination of
mean current helicity, the $\alpha_{\rm ff}$ parameter in the force-free
field extrapolation, and the gauge-invariant magnetic helicity
\citep{BF84,FA85} of the reconstructed force-free magnetic field
in the volume above an active region.
More recently, there has been growing interest in measuring magnetic
helicity {\em spectra} for selected patches at the solar surface
\citep{ZBS14,ZBS16}.
The integral of these spectra over all wavenumbers gives the mean magnetic
helicity density in the Coulomb gauge.
Furthermore, the integrated magnetic helicity spectrum weighted with
a $k^2$ factor gives the mean current helicity density based on the
vertical components of current density and magnetic field in that patch.
Unlike the magnetic helicity, it is gauge-independent, but also expected
to be more sensitive to noise resulting from the $k^2$ factor, which
amplifies the contributions from high wavenumbers $k$.
Since small-scale contributions are usually less accurate, the current
helicity is expected to be more noisy than the magnetic helicity.

Thus, an important advantage of the spectral approach is that it allows
us to filter out certain wavenumber contributions.
This is the approach adopted in the present paper.
Another advantage of the spectral approach is that it allows us to
determine the fractional helicity, which is 
a non-dimensional measure of the relative amount of magnetic helicity
that can give us a sense of the reliability or importance of a particular
measurement.
For example, one might want to discard all measurements for which the
fractional helicity is less than a certain percentage of the maximum
possible value.

Finally, we can determine the typical correlation length of the
magnetic field, which corresponds to the integral over the spectrum
weighted by $k^{-1}$ and normalized by the mean magnetic energy density.
Again, it can be used as a threshold if we are only interested in large
active regions, for example.

In a few selected cases, the measurement of magnetic helicity spectra has
revealed systematic sign changes separately for large and small scales.
An example is NOAA~11515, which emerged in the southern hemisphere, but
was found to violate the hemispheric sign rule \citep{Lim16}.
The spectral analysis showed that this sign rule violation occurred at
large scales, while the small-scale magnetic helicity still obeyed the
hemispheric sign rule. Such magnetic fields with opposite sign at large and small scales
are called bihelical \citep{YB03}. 

The bihelical nature of magnetic fields is an interesting aspect that
is actually expected based on dynamo they \citep{See96,Ji99,BB03}.
Scale-dependent sign changes of magnetic helicity have also been found
in the solar wind \citep{BSBG11} and at the solar surface \citep{Singh18}.

Here we provide an extensive study of many of the publicly available
magnetograms of {\em Hinode}, which have a pixel resolution of about $220\km$
on the Sun.
{\em Hinode}'s resolution is much better than that of the Helioseismic and
Magnetic Imager (HMI) on the {\em Solar Dynamics Observatory}, even
though the pixel size in megameters is not so different.
One must keep in mind, however, that {\em Hinode} is not a survey instrument
and that observations exist only for selected patches on the Sun.

In an associated online catalogue\footnote{
\url{http://www.nordita.org/~brandenb/projects/Hinode}}, we provide for
each of the {\em Hinode} map identification numbers the mean magnetic energy,
mean magnetic helicity, its large- and small-scale contributions, the
current helicity, fractional helicity, and the correlation length for
about 2485 maps.

\section{Method}

Following the approach of \cite{ZBS14,ZBS16} and \cite{ZB18}, we compute
the magnetic helicity spectrum as
\begin{eqnarray}
\label{EM}
H_{\rm M}(k)&=&\;\half\!\!\!\!\!\!\!\!\sum_{k_- < |{\kk}|\leq k_+} \!\!\!\!\!\!
(\tilde{A}_z\tilde{B}_z^\ast+\tilde{A}_z^\ast\tilde{B}_z),
\end{eqnarray}
where $\tilde{B}_i({\kk},t)=\int B_i({\xx},t)\,e^{i{\kk}\cdot{\xx}}d^2\xx$
is the Fourier transform of the three magnetic field components
$i=x,y,z$ of a two-dimensional Cartesian patch on the Sun
with $\xx=(x,y)$ denoting the position vector,
$\kk=(k_x,\,k_y)$ is the wavevector in the spectral plane,
$k_\pm=k\pm\delta k/2$ are the wavenumbers of an interval of width
$\delta k=2\pi/L$ around the argument $k$ of $H_{\rm M}(k)$ in \Eq{EM}
in the plane with the area $L^2$, with $L$ being the size of the
magnetogram, and
\begin{eqnarray}
\label{Az}
\tilde{A}_z=(-ik_x\tilde{B}_y+ik_y\tilde{B}_x)/k^2
\end{eqnarray}
is the vertical component of the Fourier-transformed magnetic
vector potential.

We define the total magnetic energy spectrum in the plane as
\begin{eqnarray}
\label{HM}
E_{\rm M}(k)&=&\;\half\!\!\!\!\!\!\!\!\sum_{k_- < |{\kk}|\leq k_+} \!\!\!\!\!\!
|\tilde{B}_x({\kk})|^2+|\tilde{B}_y({\kk})|^2+|\tilde{B}_z({\kk})|^2.
\end{eqnarray}
As in \cite{ZBS14}, it will be interesting to compare with
the contributions from the horizontal and vertical fields,
$E_{\rm M}^{\rm(h)}$ and $E_{\rm M}^{\rm(v)}$, respectively,
which were defined such that, if the two were equal to each
other, then both would be an approximation to the total
energy, i.e.,
$E_{\rm M}(k)\approx E_{\rm M}^{\rm(h)}\approx E_{\rm M}^{\rm(v)}$,
which requires that we define the individual contributions such that
\begin{eqnarray}
E_{\rm M}^{\rm(h)}+E_{\rm M}^{\rm(v)}=2E_{\rm M}(k).
\label{EhEvEM}
\end{eqnarray}
Specifically, we thus define them as
\begin{eqnarray}
\label{EMxy}
E_{\rm M}^{\rm(h)}(k)&=&\;\!\!\!\!\!\!\!\!\sum_{k_- < |{\kk}|\leq k_+} \!\!\!\!\!\!
|\tilde{B}_x({\kk})|^2+|\tilde{B}_y({\kk})|^2,\\
\label{EMz}
E_{\rm M}^{\rm(h)}(k)&=&\;\!\!\!\!\!\!\!\!\sum_{k_- < |{\kk}|\leq k_+} \!\!\!\!\!\!
|\tilde{B}_z({\kk})|^2,\\
\end{eqnarray}
i.e., without the $1/2$ factor in \Eq{HM}, so that \Eq{EhEvEM} is obeyed.

With our approach, we obtain the mean magnetic energy and helicity densities
in the plane as
\begin{eqnarray}
{\cal E}_{\rm M}=\int_0^\infty E_{\rm M}(k)\,dk,\quad
{\cal H}_{\rm M}=\int_0^\infty H_{\rm M}(k)\,dk.
\end{eqnarray}
Since most of the magnetic energy and helicity in the plane comes from
the active region and not the space around it, it makes sense to multiply
${\cal E}_{\rm M}$ and ${\cal H}_{\rm M}$ by the size of the patch, $L^2$.
Furthermore, to facilitate comparison with results in the literature,
\cite{ZBS14} chose to compute energy and helicity over an arbitrarily
defined volume of height $L_z=100\Mm$ above the active region.
We adopt here the same approach and thus quote the values of
\begin{eqnarray}
e_{\rm M}={\cal E}_{\rm M}L^2L_z\quad\mbox{and}\quad
h_{\rm M}={\cal H}_{\rm M}L^2L_z.
\end{eqnarray}

We also determine the large-scale (LS) and small-scale (SS) contributions
to the magnetic helicity by defining
\begin{eqnarray}
{\cal H}_{\rm M}^{\rm LS}=\int_0^{k_{\rm LS}} H_{\rm M}(k)\,dk\quad
{\cal H}_{\rm M}^{\rm SS}=\int_{k_{\rm SS}}^\infty H_{\rm M}(k)\,dk,
\end{eqnarray}
where we chose $k_{\rm LS}=0.4\Mm^{-1}$ and $k_{\rm SS}=3\Mm^{-1}$ as
the limiting wavenumbers marking the end of the LS range and the beginning
of the SS range, respectively.
This choice can be motivated by inspecting several examples of spectra that
show similar signs of spectral magnetic helicity in the ranges
$k<k_{\rm LS}$ and $k>k_{\rm SS}$; see the aforementioned website for the
online catalogue.

As alluded to above, we also compute the correlation length of the magnetic
field, which is defined as

\begin{eqnarray}
\ell_{\rm M}=\int_0^\infty k^{-1} E_{\rm M}(k)\,dk\left/
\int_0^\infty E_{\rm M}(k)\,dk.\right.
\end{eqnarray}
This allows us to compute the fractional helicity as
\begin{eqnarray}
r_{\rm M}={\cal H}_{\rm M}/2\ell_{\rm M}{\cal E}_{\rm M}.
\end{eqnarray}
The value of $r_{\rm M}$ lies in the range $-1\leq\ell_{\rm M}\leq1$.

\section{Observational Data}
We use high resolution and high-sensitivity vector magnetograms provided
as level-2 data products by the Milne-Eddington inversion pipeline
MERLIN at HAO/CSAC (DOI:10.5065/D6JH3J8D).
These vector magnetograms are deduced from the spectropolarimetric
scans of solar magnetic regions by the {\em Hinode} Solar Optical
Telescope/Spectro-Polarimeter (SOT/SP) instrument \citep{Tsun08} that
has a diffraction limited field-of-view of up to $328''\times164''$
and an angular resolution of $0.3''$.
More details about the {\it Hinode} SOT/SP instrument and the calibration
of data can be found in \cite{Lites13a} and \cite{Lites13b}.
The level-2 data products consist of area scans of a variety of target
regions such as active regions, quiet sun, polar regions, and repeated
small region scans for time evolution studies.
For our study we down-selected these data to include only active regions
and pores. 
The majority (73\%) of the selected data are sunspots or active regions
with fully formed penumbrae, while the rest (27\%) are pores without penumbrae.
The level-2 vector magnetograms were resolved for the 180$^\circ$
azimuth ambiguity using the method described in \cite{RA14}.

We selected the data based on the following criteria:
\begin{itemize}\itemsep-.5mm
\item
The observed region should be inside the heliocentric angle range of 0
to 30 degrees. This is done to avoid perspective effects and the need to
do a heliographic coordinate transformation of the vector magnetograms.
\item
The field-of-view of the observed region should be at least $96''$
in either direction. This is done to avoid partial/incomplete scans
of active regions.
\item
The total area occupied by dark umbra or pores in the observed region should
be greater than $(10'')^2$, or 900 pixels. This is done to avoid
selecting very small-sized pores. The umbral area is computed from the
continuum intensity map by first removing the limb darkening function
and then normalizing the intensity to median value in non-magnetic
pixels. Pixels with normalized intensity less than or equal to 0.55
are treated as umbra or pore.
\item
The data that satisfies the above criteria sometimes include undesired
characteristics such as repeated small area scans for time evolution
studies of sunspots or active region scans with missing scan lines, bad
columns, or partial scans. Thus, as a final criterion, the data is displayed
and manually rejected if these undesired characteristics are present.
\end{itemize}

\begin{figure}[t!]\begin{center}
\includegraphics[width=\columnwidth]{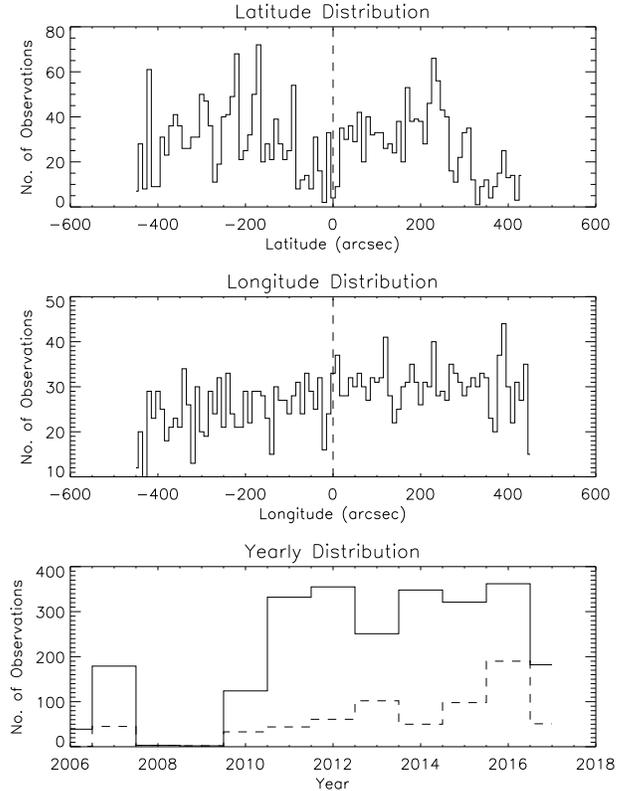}
\end{center}\caption{Distribution of {\em Hinode} observations selected for
this study in latitude, longitude and year-wise shown in top, middle and
bottom panel, respectively.
The solid line in the histogram in the bottom panel represents
all observations selected for analysis, while the dashed line shows
the fraction of those containing pores.
}\label{observ_distr}\end{figure}

The distribution of latitude, longitude and year of the selected
observations is shown in \FFig{observ_distr}.
The yearly distribution is found to be relatively uniform, except for the time of solar
minimum during 2008 and 2009.

{\em Hinode} observes the target regions with either normal mode
($0.16''/{\rm pixel}$ sampling) or fast mode
($0.32''/{\rm pixel}$ sampling).
In our selected dataset, both modes exist.
We convert normal mode scans in our dataset to $0.32''/{\rm pixel}$,
so that all maps have the same spatial sampling.
Further, in our calculations we always use $512\times512$ pixels in the
region-of-interest (ROI).
If the original data is larger, we extract $512\times512$ ROI centered
pixels around the active region or pore.
On the other hand, if the original data is smaller, we embed the
observed region in the center of a $512\times512$ array with zero padding
in adjacent missing pixels.

Finally, we create a database of helicity parameters for each scan,
which is uniquely identified by the {\em Hinode} MapID; see the aforementioned
website, which also contains spectra for each map.

\section{Results}
We have processed 2485 vector magnetograms over the solar disc for the
years 2006 through 2017, covering in some cases the entire evolution of
an active region as it passes the solar disc. 
Of these magnetograms, 680 correspond to pores (dark regions without
penumbrae) and 1805 to fully developed sunspots and active regions with fully
developed penumbrae.
There can be significant temporal variations of helicity, which are
sometimes associated with the development of flares and coronal mass
ejections.

\begin{table}[b!]\caption{
Percent of Active Regions following Hemispheric Rule 
}\vspace{12pt}\centerline{\begin{tabular}{lccccr}
\hline
\\
$\!\!\!\!$Hemisph.$\!\!$ & $h_{\rm M} [\%]$& $h_{\rm M}^{\rm LS} [\%]$& $h_{\rm M}^{\rm SS} [\%]$ & ${\cal H}_{\rm C}[\%]$ & $r_{\rm M} [\%]$ \\
\hline
North  & 62$\pm$3 & 61$\pm$3 & 24$\pm$2 & 47$\pm$3 & 62$\pm$3  \\
South  & 59$\pm$3 & 58$\pm$3 & 82$\pm$2 & 73$\pm$2 & 59$\pm$3  \\ 
\hline
\label{ns_distr}\end{tabular}}
\end{table}

\begin{figure*}[t!]\begin{center}
\includegraphics[width=5.5in]{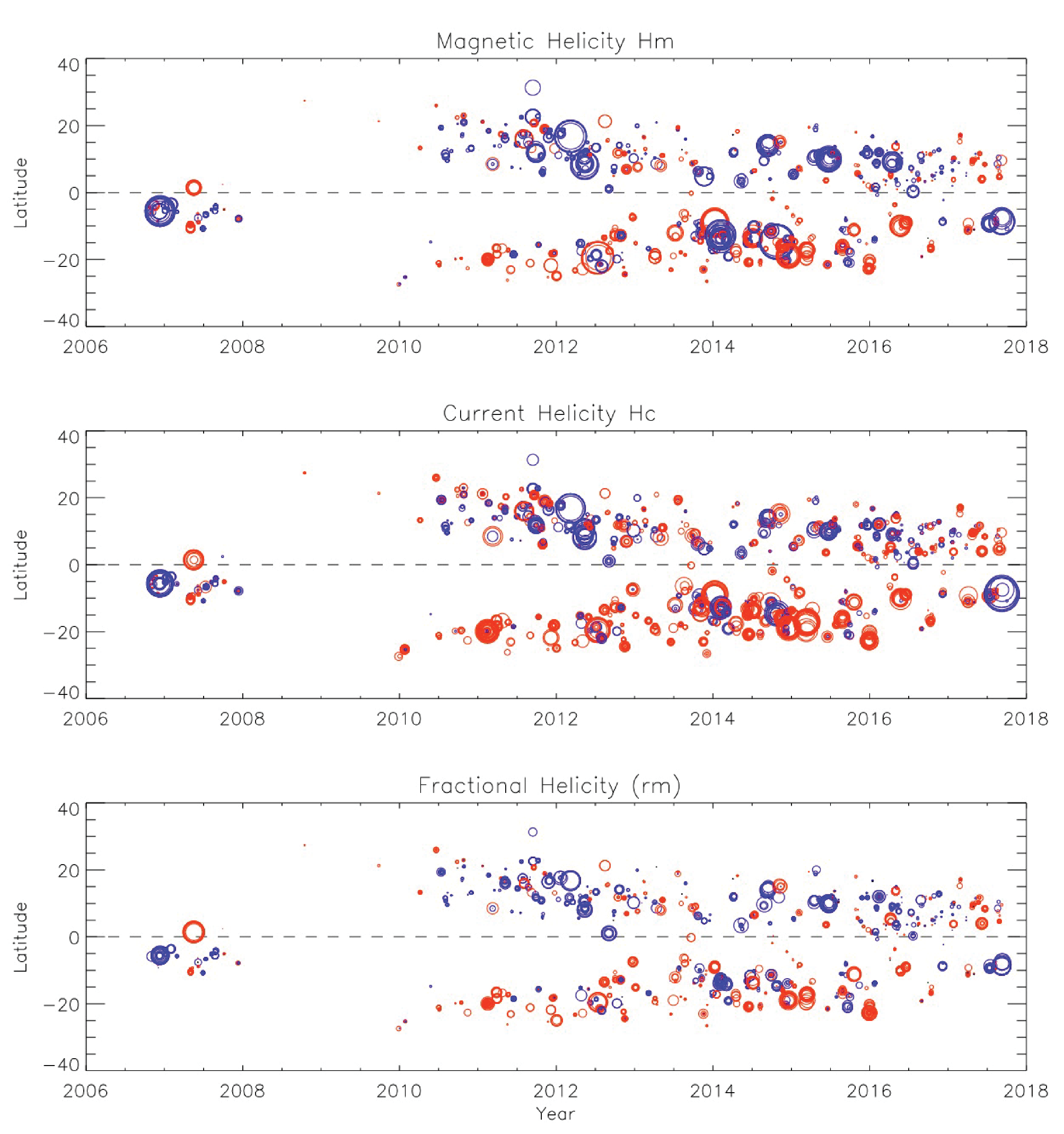}
\end{center}\caption{
Time-latitude distribution of helicity parameters.  Top panel shows the
distribution of magnetic helicity density (${\cal H}_{\rm M}$), middle
panel the current helicity density (${\cal H}_{\rm C}$), and the bottom
panel shows the fractional helicity ($r_{\rm M}$).
The blue (red) circles represents negative (positive) sign of these
parameters.
In the top two panels, we scaled the diameter of the circles to the square
root of the normalized amplitudes, since the values vary over a large range.
In the bottom panel, the values are fractional quantities between $\pm1$,
so the radii of the circles are just scaled to the $r_{\rm M}$ value.
The biggest circle in each of the three plots corresponds, respectively, to $37.8\times 10^{42} \Mx^2$,
$875.2\times 10^{24} \G^2 \cm^{-1}$, and 0.61.
}\label{butterfly}
\end{figure*}

\begin{table*}[htb]\caption{
Summary of Data for NOAA~10930.
}\vspace{12pt}\centerline{\begin{tabular}{crrrrrrrrrrrrc}
y & m & d & h:m & $e_{\rm M}$ & $h_{\rm M}$ &
$h_{\rm M}^{\rm LS}$ & $h_{\rm M}^{\rm SS}$ &
${\cal H}_{\rm C}$ & $r_{\rm M}$ & $\ell_{\rm M}$ & $\lambda$ & ${\cal L}$ & MapID \\
\hline
2006&12& 9&10:00&$10.8$&$ -8.8$&$ -8.8$&$ -2.3$&$ -202$&$-0.22$&$ 7.56$&$ -5.7$&$-26.9$&$ 30107$\\
2006&12& 9&11:20&$11.0$&$ -8.7$&$ -8.7$&$ -5.5$&$ -179$&$-0.21$&$ 7.57$&$ -5.7$&$-26.3$&$ 30108$\\
2006&12& 9&12:40&$11.1$&$ -8.6$&$ -8.5$&$  2.7$&$ -157$&$-0.20$&$ 7.64$&$ -5.7$&$-25.5$&$ 30109$\\
2006&12& 9&14:00&$11.2$&$ -8.0$&$ -8.0$&$  4.8$&$ -140$&$-0.19$&$ 7.60$&$ -5.7$&$-24.8$&$ 30110$\\
2006&12& 9&17:10&$11.3$&$ -8.1$&$ -8.1$&$ 10.7$&$ -117$&$-0.19$&$ 7.61$&$ -5.7$&$-23.0$&$ 30111$\\
2006&12& 9&22:00&$11.8$&$ -8.1$&$ -8.0$&$ 11.9$&$ -128$&$-0.18$&$ 7.61$&$ -5.7$&$-20.4$&$ 30112$\\
2006&12&10& 1:00&$12.1$&$ -8.1$&$ -8.0$&$  4.9$&$ -144$&$-0.17$&$ 7.62$&$ -5.7$&$-18.7$&$ 30113$\\
2006&12&10&10:55&$13.7$&$ -8.2$&$ -8.2$&$  6.4$&$ -197$&$-0.16$&$ 7.44$&$ -5.7$&$-12.6$&$ 30114$\\
2006&12&10&21:00&$14.4$&$-14.7$&$-14.5$&$ -4.4$&$ -489$&$-0.27$&$ 7.66$&$ -5.7$&$ -7.6$&$ 30115$\\
2006&12&11& 3:10&$15.2$&$-20.9$&$-20.7$&$-14.7$&$ -720$&$-0.35$&$ 7.73$&$ -5.7$&$ -4.1$&$ 30116$\\
2006&12&11& 8:00&$15.5$&$-24.3$&$-24.0$&$-11.6$&$ -806$&$-0.40$&$ 7.71$&$ -5.7$&$ -1.5$&$ 30117$\\
2006&12&11&11:10&$15.6$&$-26.8$&$-26.6$&$-12.1$&$ -802$&$-0.44$&$ 7.84$&$ -5.7$&$  0.4$&$ 30118$\\
2006&12&11&13:10&$15.7$&$-28.3$&$-28.0$&$-10.6$&$ -848$&$-0.46$&$ 7.81$&$ -5.7$&$  1.9$&$ 30119$\\
2006&12&11&17:00&$16.2$&$-29.9$&$-29.7$&$ -6.4$&$ -850$&$-0.47$&$ 7.87$&$ -5.7$&$  3.5$&$ 30120$\\
2006&12&11&20:00&$16.2$&$-30.7$&$-30.4$&$-10.8$&$ -862$&$-0.48$&$ 7.86$&$ -5.7$&$  5.2$&$ 30121$\\
2006&12&11&23:10&$16.2$&$-32.4$&$-32.2$&$-12.7$&$ -822$&$-0.50$&$ 7.92$&$ -5.7$&$  7.0$&$ 30122$\\
2006&12&12& 3:50&$16.7$&$-33.0$&$-32.8$&$-16.7$&$ -892$&$-0.51$&$ 7.78$&$ -5.7$&$  9.6$&$ 30123$\\
2006&12&12&10:10&$16.3$&$-33.0$&$-32.8$&$-19.2$&$ -875$&$-0.52$&$ 7.80$&$ -5.5$&$ 13.6$&$ 30124$\\
2006&12&12&15:30&$15.8$&$-29.7$&$-29.5$&$-17.1$&$ -781$&$-0.48$&$ 7.77$&$ -5.7$&$ 16.0$&$ 30125$\\
2006&12&12&17:40&$15.6$&$-28.1$&$-27.9$&$-17.5$&$ -696$&$-0.46$&$ 7.81$&$ -5.7$&$ 17.2$&$ 30126$\\
2006&12&12&20:30&$15.0$&$-25.8$&$-25.7$&$-11.2$&$ -596$&$-0.43$&$ 7.88$&$ -5.7$&$ 18.8$&$ 30127$\\
2006&12&13& 4:30&$14.8$&$-29.1$&$-29.0$&$-14.1$&$ -638$&$-0.49$&$ 7.91$&$ -5.7$&$ 23.2$&$ 30128$\\
2006&12&13& 7:50&$14.2$&$-28.3$&$-28.2$&$ -8.7$&$ -618$&$-0.50$&$ 7.96$&$ -5.7$&$ 25.1$&$ 30129$\\
\label{AR10930}\end{tabular}}
\tablenotemark{
$e_{\rm M}$ is in $10^{32}\erg$, $h_{\rm M}$ and $h_{\rm M}^{\rm LS}$
are in $10^{42}\Mx^2$, $h_{\rm M}^{\rm SS}$ is in $10^{38}\Mx^2$, ${\cal H}_{\rm C}$ is
in $10^{24}\G^2\cm^{-1}$, $r_{\rm M}$ is dimensionless, $\ell_{\rm M}$
is in Mm, and $\lambda$ and ${\cal L}$ are in degrees.}
\end{table*}

\subsection{Time-Latitude Distribution of Helicity}
{\em Hinode} data selected here span almost a solar cycle, so we first look at
the distribution of helicity sign and magnitude with time and latitude
during the end of cycle 23 and most of the cycle 24.
In \Fig{butterfly}, the distributions of $H_{\rm M}$, $H_{\rm C}$, and
$r_{\rm M}$ are given.
The negative (positive) sign of these parameters are represented by
blue (red) color.
As is found in many previous studies, the statistical trend of negative
(positive) sign in the northern (southern) hemisphere is present.
The relative amplitude of these parameters is represented by radius
of the circle symbol in \FFig{butterfly}.
We summarize the hemispheric statistics of these parameters in
\Tab{ns_distr} with 95\% confidence intervals.
It is seen that hemispheric bias is present and is significant in
$h_{\rm M}$, $h_{\rm M}^{\rm LS}$, and $r_{\rm M}$ in both hemispheres.
While for current helicity (${\cal H}_{\rm C}$) the bias is weak in the
north, it is found to be strong in the south.

For small scales, $h_{\rm M}^{\rm SS}$ shows a peculiar result in that the sign is
predominantly positive in both north (75\%) and south (83\%).
This is perhaps because, as seen in \Tab{AR10930}, most of the helicity
is accounted for by the large-scale component.
Typically, for all of the data the amplitude of small-scale helicity is
about $10^{4}$ times smaller than the large-scale helicity.
Thus, most of the contribution must come from the large-scale part.
There is good agreement between the $h_{\rm M}$, $h_{\rm M}^{\rm LS}$, and
$r_{\rm M}$ statistics.

\subsection{Latitudinal Dependence}

\begin{figure}[t!]\begin{center}
\includegraphics[width=\columnwidth]{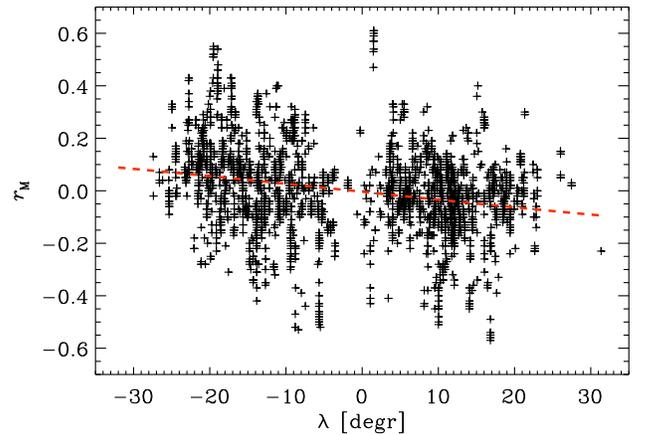}
\end{center}\caption{
Dependence of fractional magnetic helicity on latitude.
}\label{prM_lat}\end{figure}

The dependence of the fractional magnetic helicity on Carrington latitude
$\lambda$ is shown in \Fig{prM_lat}.
This relation is extremely noisy, although there is still a clear negative
correlation with $\lambda$.
Specifically, we find
\begin{eqnarray}
r_{\rm M}(\lambda)=-0.004-0.17\sin\lambda.
\end{eqnarray}
The dependences of ${\cal H}_{\rm M}(t)$ and ${\cal H}_{\rm C}(t)$
on latitude (not shown) are even more noisy, but they also show
negative correlations:
\begin{eqnarray}
h_{\rm M}(\lambda)=-0.56-3.2\sin\lambda\quad[\G^2\Mm^{4}],
\end{eqnarray}
\begin{eqnarray}
{\cal H}_{\rm C}(\lambda)=18-170\sin\lambda\quad[\G^2\km^{-1}].
\end{eqnarray}

Previous statistical studies of the latitudinal variation of
helicity parameters derived from other observations also show similar
scatter \citep{HS04,XG07,SG13}.
It is also observed that the hemispheric helicity trend varies during
the solar cycle.
\cite{HS05} studied the annual variation of the helicity trend and found
that the hemispheric sign preference is more likely to be present during
the maximum phase rather than during the minimum phase of the solar cycle.
There are also signatures that the hemispheric trend may reverse during
the declining phase of the solar cycle \citep{BAZ00,HQZ10,PPLK19}.

During the activity minimum phase, the statistics are generally poorer
due to a small number of active regions than during the maximum phase.
Here we only present the latitudinal distribution of all observations
during a cycle, which is a better statistic and is less prone to biases
due to selection effects such as size and number of active regions during
a single year.
Synoptic full-disk vector magnetograms are generally better suited for
the study of the time variation of the hemispheric helicity trend.

The amount of scatter is rather significant and seems to support a similar
trend from earlier findings suggesting that at higher resolution, the
general hemispheric sign rule deteriorates; cf.\ the  earlier findings
by \cite{BZ98} and \cite{PCM95}, where the latter showed much stronger
scatter than the former.
Similarly, using low resolution vector synoptic maps, \cite{SG13} found
a weak hemispheric trend with smaller scatter in the current helicity
density.

\begin{figure}[t!]\begin{center}
\includegraphics[width=\columnwidth]{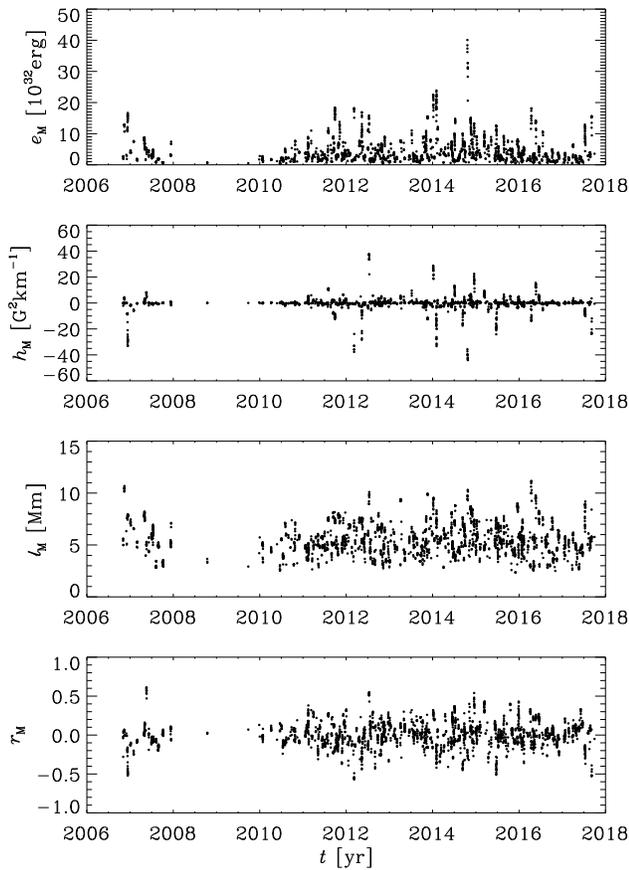}
\end{center}\caption{
Temporal variation of $e_{\rm M}(t)$, $h_{\rm M}(t)$,
$\ell_{\rm M}(t)$, and $r_{\rm M}(t)$ for all 2485 maps.
}\label{pEHt}\end{figure}

\subsection{Time Dependence}
There is a general hemispheric preference with most of the active regions
having negative magnetic helicity in the north and positive in the south.
However, there can also be significant departures from this hemispheric
preference.
\FFig{pEHt} shows the evolution of $e_{\rm M}(t)$,
$h_{\rm M}(t)$, $\ell_{\rm M}(t)$, and $r_{\rm M}(t)$
for all 2485 maps, regardless of position or selection effects
arising from the fact that particularly interesting active
regions have been observed repeatedly.
One clearly sees overall enhanced activity during solar maximum
around 2014 and only very few measurements during solar minimum
around 2008 and 2009.
While $e_{\rm M}(t)$ does show some intense spikes of activity
on a timescale of 1--2 years, the spikes in $h_{\rm M}(t)$
are even more extreme.
This is reminiscent of earlier findings using a related method applied to
synoptic vector magnetograms \citep{BPS17}.
On the other hand, $r_{\rm M}(t)$ seems to be now less spiky
than what has been found from the synoptic vector magnetograms.
This difference can well be caused by the aforementioned selection
effects resulting from the fact that particularly interesting regions
have been observed more frequently.

The overall variation of $\ell_{\rm M}(t)$ is rather small and the values
are around $6\Mm$ both during minimum and maximum.
Similar values have also been found with both HMI and the Huairou Solar
Observing Station \citep{ZBS16}.
This value of $\ell_{\rm M}$ is significantly smaller than what
has been found using the synoptic vector magnetograms from HMI, where
$\ell_{\rm M}$ was found to fluctuate around $20\Mm$, or from the synoptic
vector magnetograms from SOLIS, where $\ell_{\rm M}$ was found to fluctuate
around $15\Mm$ \citep{Singh18}.

As already emphasized by \cite{Singh18}, the numerical value of
$\ell_{\rm M}$ must not be interpreted as a physically identifiable
length scale.
In fact, since it is defined as a weighted inverse wavenumber, it might
make sense to identify $2\pi\ell_{\rm M}$ with a physically relevant
length scale.

The fact that $\ell_{\rm M}$ is about three times larger when
it is determined from the synoptic maps is interesting and has
not previously been noticed.
This may indicate that a synoptic magnetogram is different from
an actual magnetogram.
It could be caused by an anisotropy resulting from the assembly
of different magnetograms in the longitudinal direction.
This aspect is worth revisiting in future.

\begin{figure}[t!]\begin{center}
\includegraphics[width=.95\columnwidth]{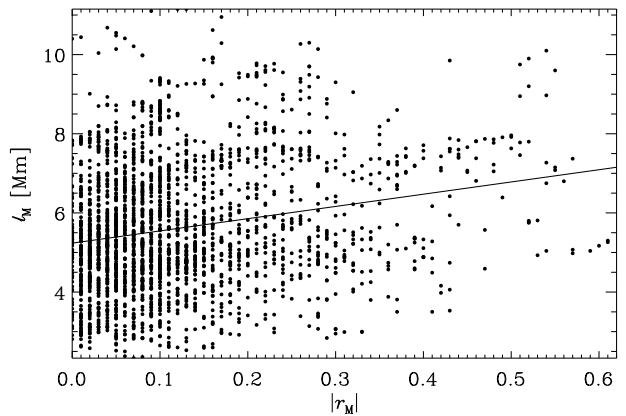}
\end{center}\caption{
Scatter plot between $r_{\rm M}(t)$ and $\ell_{\rm M}(t)$.
}\label{pell_rm}\end{figure}

\begin{figure}[t!]\begin{center}
\includegraphics[width=\columnwidth]{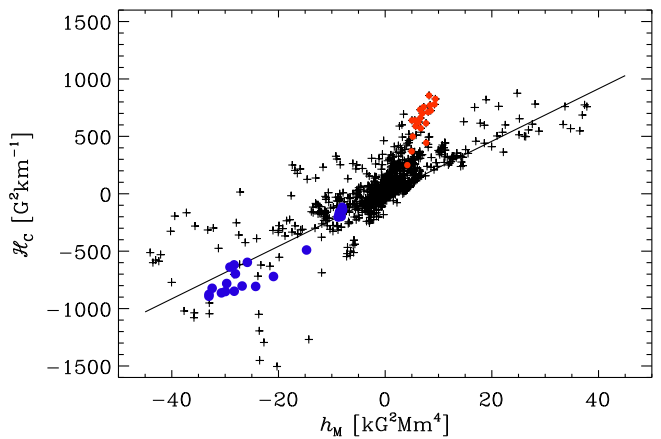}
\end{center}\caption{
Scatter plot showing the dependence of current helicity on magnetic helicity.
The blue (red) symbols show the data for NOAA~10930 (NOAA~12297) only.
}
\label{pHC_HM}
\end{figure}

Earlier work by \cite{ZBS16} showed that $\ell_{\rm M}(t)$
displays a clear modulation with the solar cycle, where
$\ell_{\rm M}(t)$ varied between $6\Mm$ during solar minimum
and $8\Mm$ during solar maximum.
No such clear variation can be seen from our current data.
Nevertheless, looking at a scatter plot between $r_{\rm M}(t)$ and $\ell_{\rm M}(t)$
does suggest a positive, albeit very noisy correlation between the two;
see \Fig{pell_rm}.

In this connection it is useful to recall the findings of \cite{YZ12}
and \cite{ZY13} that the quiet sun contributes much less to the cyclic
variation than active regions.
This could explain the relatively small variation of $\ell_{\rm M}(t)$
in our data, because with {\em Hinode} we expect a stronger and more accurate
contribution from the quiet sun than for the Huairou Solar Observing
Station.

\begin{table*}[htb]\caption{
Summary of Data for NOAA~12297.
}\vspace{12pt}\centerline{\begin{tabular}{crrrrrrrrrrrrc}
y & m & d & h:m & $e_{\rm M}$ & $h_{\rm M}$ &
$h_{\rm M}^{\rm LS}$ & $h_{\rm M}^{\rm SS}$ &
${\cal H}_{\rm C}$ & $r_{\rm M}$ & $\ell_{\rm M}$ & $\lambda$ & ${\cal L}$ & MapID \\
\hline
2015& 3& 9&20:48&$ 5.9$&$  5.0$&$  4.7$&$ 10.1$&$  370$&$ 0.36$&$ 4.66$&$-20.0$&$-21.6$&$115255$\\
2015& 3&10& 7:45&$ 6.7$&$  6.2$&$  5.8$&$ 18.3$&$  626$&$ 0.41$&$ 4.49$&$-19.6$&$-18.7$&$115262$\\
2015& 3&10&11:04&$ 6.4$&$  5.7$&$  5.3$&$ 20.6$&$  588$&$ 0.39$&$ 4.57$&$-18.1$&$-19.1$&$115264$\\
2015& 3&11& 3:15&$ 7.3$&$  6.6$&$  6.2$&$ 24.5$&$  660$&$ 0.39$&$ 4.60$&$-17.1$&$-28.2$&$115267$\\
2015& 3&11& 8:10&$ 7.7$&$  6.7$&$  6.4$&$ 17.9$&$  571$&$ 0.36$&$ 4.74$&$-17.1$&$-25.5$&$115268$\\
2015& 3&11&22:01&$ 9.2$&$  8.1$&$  7.8$&$ 14.6$&$  712$&$ 0.35$&$ 4.95$&$-17.1$&$-17.8$&$115271$\\
2015& 3&11&22:35&$ 8.7$&$  8.5$&$  8.2$&$ 16.3$&$  724$&$ 0.40$&$ 4.89$&$-17.1$&$-17.5$&$115272$\\
2015& 3&12& 3:22&$ 9.4$&$  9.4$&$  9.1$&$ 23.2$&$  826$&$ 0.41$&$ 4.85$&$-17.2$&$-14.8$&$115273$\\
2015& 3&12& 4:43&$ 9.5$&$  9.2$&$  9.0$&$ 15.0$&$  778$&$ 0.40$&$ 4.88$&$-17.2$&$-14.1$&$115274$\\
2015& 3&12&10:37&$10.7$&$  8.2$&$  7.8$&$ 25.9$&$  856$&$ 0.31$&$ 4.90$&$-17.2$&$-10.8$&$115275$\\
2015& 3&12&13:52&$ 9.0$&$  8.4$&$  8.0$&$ 23.8$&$  771$&$ 0.39$&$ 4.78$&$-17.2$&$ -4.6$&$115276$\\
2015& 3&12&15:50&$10.0$&$  7.2$&$  6.9$&$ 16.2$&$  756$&$ 0.29$&$ 4.89$&$-17.2$&$ -4.6$&$115277$\\
2015& 3&12&21:00&$ 9.5$&$  7.0$&$  6.6$&$ 15.1$&$  701$&$ 0.30$&$ 4.83$&$-17.2$&$ -5.1$&$115278$\\
2015& 3&12&21:48&$ 7.5$&$  7.7$&$  7.3$&$ 14.5$&$  614$&$ 0.43$&$ 4.74$&$-17.2$&$ -4.7$&$115279$\\
2015& 3&13& 3:01&$ 8.6$&$  6.6$&$  6.2$&$ 21.4$&$  732$&$ 0.32$&$ 4.68$&$-17.2$&$ -1.8$&$115280$\\
2015& 3&13&10:30&$ 7.5$&$  5.6$&$  5.2$&$ 17.0$&$  639$&$ 0.32$&$ 4.54$&$-17.2$&$  2.4$&$115282$\\
2015& 3&13&20:00&$ 7.2$&$  5.0$&$  4.6$&$ 18.6$&$  639$&$ 0.30$&$ 4.61$&$-17.2$&$  5.0$&$115283$\\
2015& 3&14& 1:50&$ 6.9$&$  5.2$&$  4.9$&$ 11.2$&$  498$&$ 0.31$&$ 4.78$&$-17.2$&$ 10.8$&$115285$\\
2015& 3&15& 9:30&$ 5.9$&$  7.7$&$  7.5$&$ 11.1$&$  444$&$ 0.43$&$ 5.94$&$-17.3$&$ 28.1$&$115292$\\
2015& 3&16&23:00&$ 3.8$&$  4.2$&$  4.1$&$  1.5$&$  250$&$ 0.41$&$ 5.28$&$-19.0$&$ 24.5$&$115302$\\
\label{AR12297}\end{tabular}}
\tablenotemark{
$e_{\rm M}$ is in $10^{32}\erg$, $h_{\rm M}$ and $h_{\rm M}^{\rm LS}$
are in $10^{42}\Mx^2$, $h_{\rm M}^{\rm SS}$ is in $10^{38}\Mx^2$, ${\cal H}_{\rm C}$ is
in $10^{24}\G^2\cm^{-1}$, $r_{\rm M}$ is dimensionless, $\ell_{\rm M}$
is in Mm, and $\lambda$ and ${\cal L}$ are in degrees.}
\end{table*}

\begin{figure*}[t!]\begin{center}
\includegraphics[width=5.8in]{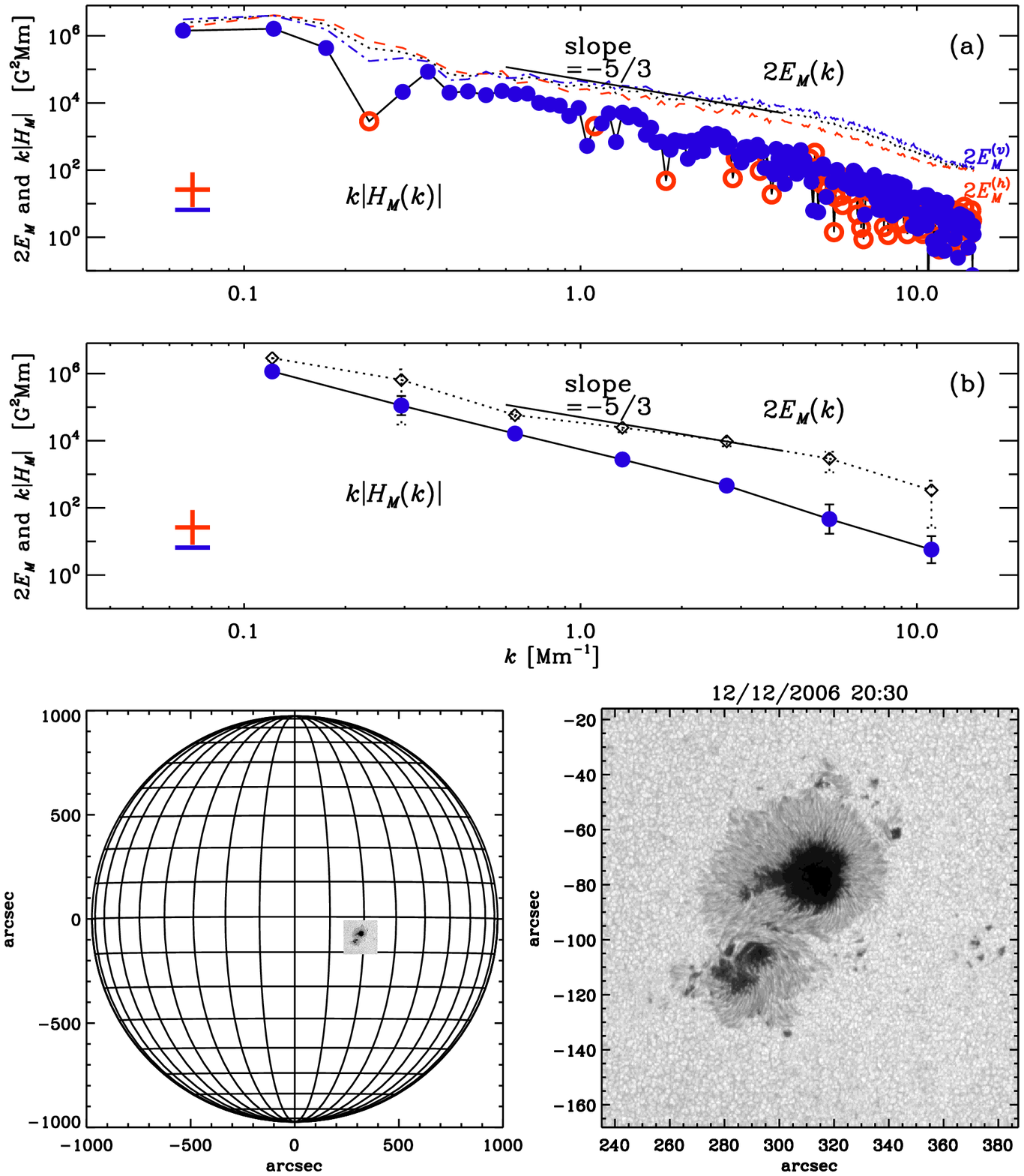}
\end{center}\caption{
(a) 2$E_{\rm M}(k)$ (dotted line) and $k|H_{\rm M}(k)|$ (solid line) for NOAA~10930
at 20:30 UT on 2006 December 12.
Positive (negative) values for $H_{\rm M}(k)$ are indicated by open (closed)
symbols, respectively.
$2E^{(h)}_{\rm M}(k)$ (red, dashed) and $2E^(v)_{\rm M}(k)$ (blue, dash-dotted) are
shown for comparison.
(b) Same as (a), but the magnetic helicity is averaged over broad
logarithmically spaced wavenumber bins.
}\label{sunspot}\end{figure*}

\subsection{Relation with Current Helicity}

In homogeneous turbulence, there is a relation between the magnetic
helicity spectrum and the current helicity spectrum such that
$H_{\rm C}(k)=k^2H_{\rm M}(k)$.
There is no such clear relationship between magnetic and current
helicity in physical space, although the two might still be related
to each other by the square of a length scale.

In \Fig{pHC_HM} we show the dependence of magnetic helicity
on current helicity as a scatter plot.
We see a positive dependence with a slope $23\times10^{-6}\Mm^{-5}$.
Adopting again our reference volume of $V=(100\Mm)^3$ used in our
calculations of $H_{\rm M}(k)$, we find $k^2=23\Mm^{-2}$,
i.e., $k=4.8\Mm^{-1}$ or $2\pi/k=1.3\Mm$.
This corresponds to the scale of granulation.
Such an association between the typical scale of current helicity
patterns and granulation has not previously been possible to make.

\FFig{pHC_HM} seems to show evidence of a separate group of points
with a slightly steeper correlation.
This group of points comprises solely those belonging to NOAA~12297,
as is demonstrated by the red symbols in that figure.
This is an active region at $-17\degr$ latitude, which has a rather
large value of $r_{\rm M}$ of $0.3$ to $0.4$; see \Tab{AR12297}.
However, the more important exception here is that for NOAA~12297,
$h_{\rm M}$ is rather small ($5$--$9\kG^2\Mm^4$) in comparison with
NOAA~10930, where it reaches values of around $20$--$30\kG^2\Mm^4$.

\subsection{Case Study: NOAA~10930}

The tabulated values of various parameters for the well studied
active region NOAA~10930 during December 2006 where already shown in
\Tab{AR10930}.
An example of helicity and energy spectra for this active region during
12 December 2006 at 20:30 UT is shown in \FFig{sunspot}.
We find that the magnetic helicity for this active region is negative
during 9 through 13 December 2006.
This sign is opposite to the expectation from the hemispheric helicity rule.
The negative sign is seen in all helicity indicators in \Tab{AR10930},
except for the small-scale magnetic helicity, $h_{\rm M}^{\rm SS}$, 
during the early stage of the active region evolution from 9 December 12:40 UT
to 10 December 10:55 UT, after which $h_{\rm M}^{\rm SS}$ is negative. 
It is interesting to note that, during the early stages of flux emergence,
this active region has a similar pattern of magnetic helicity as NOAA~11515
\citep{Lim16}, i.e., with opposite signs at large and small scales.

The active region was flare productive and led to 3 M-class, 3 X-class,
and several C-class flares.
Many authors have reported strong rotating motion in one of the spots
in this group \citep{Yan09}.
Using the three-dimensional nonlinear force-free field (NLFFF)
extrapolation method, \cite{Park10} computed the relative coronal magnetic
helicity for this active region to be about $-4.3\times 10^{43}$ Mx$^2$
just before an X3.4 flare on 13 December 2006.
In comparison, our magnetic helicity estimate, $h_{\rm M}$, for this
time is about $-2.6\times 10^{43}$ Mx$^2$.
We notice that the time of peak helicity in this active region from
\cite{Park10} and our estimates is the same, i.e., around 3:50 to 10:10
UT on 12 December 2006.
\cite{Park10} suggest that the evolution of helical structures of opposite
sign to the active region dominant helicity sign led to flaring activity in this
active region.

\cite{Ravi11} studied the evolution of net electric currents in this
active region and found that the dominant current in the two opposite
magnetic polarities is of opposite sign, i.e., upward electric current
in one polarity and downward in another.
Further, they found that the net current in both polarities decreases
before the flares and attributed this decrease to an increase in the
non-dominant oppositely signed currents in each polarity.

The helicity spectra in the top panel of \FFig{sunspot} do show
helicity of both signs in general, but the dominant sign is negative
when averaged over logarithmically spaced wavenumber bins.
The evolution of such helicity spectra at different scales and their
relationship with flaring and/or eruptive activity could be insightful.
We defer such study in flaring regions to a future work.

\subsection{Spectral energy for vertical and horizontal fields}

It is instructive to look at magnetic energy spectra separately for
horizontal and vertical (or radial) magnetic fields.
The two are remarkably similar at all wavenumbers; see \Fig{sunspot}.
This is rather different from the earlier results by
\cite{ZBS14}\footnote{We use here the opportunity to correct a labeling
error in their Figure~2, where the energies of vertical and horizontal
fields should have been swapped.}, who found significant
departures at small scales, where the horizontal contribution was
found to exceed the vertical one by a factor of about three.

The reason for the small-scale excess of horizontal over vertical field
strengths may well be physical, but it is striking that with the higher
resolution of {\em Hinode}, the two spectra track each other much better than
with HMI.
Looking at \Fig{sunspot}, the two spectra agree nearly
perfectly up to $k=10\Mm^{-1}$, which corresponds to a scale of
$(2\pi/10)\Mm\approx600\km$.
This leads us to expect that with even higher resolution such as that
of Daniel K.\ Inouye Solar Telescope, we may continue to see the two
spectra tracking each other up to larger wavenumbers at higher resolution.
It also suggests that, if we regard the wavenumber where the spectra
depart from each other as the resolution limit, this limit is poorer
than previously anticipated.
Indeed, with HMI, we see departures already at scales of around $2\Mm$.
Much of this departure is possibly caused by intrinsic artifacts outside
the strong-field regions in the HMI magnetograms.
Those should be investigated in subsequent analyzes.

\section{Conclusions}

The purpose of the present work was to use {\em Hinode} data to provide a
comprehensive survey of spectral magnetic helicity.
The data turn out to be of considerably higher quality than those used
in earlier analyzes of HMI and SOLIS data.
This became evident when comparing magnetic energy spectra separately
for vertical and horizontal magnetic field components.
Unlike earlier work using HMI, which showed significant departure
between the two at $k=3\Mm^{-1}$ \citep{ZBS14}, the present analysis
shows the two spectra tracking each other up $k=10\Mm^{-1}$.

The correlation length $\ell_{\rm M}$, on the other hand, appears to
be rather similar between current and earlier analyzes.
However, there are differences in comparison with similar results using
synoptic magnetograms.
Those differences are tentatively associated with the anisotropy resulting
from combining magnetograms of different times into a new map.

A major surprise arising from our work is the poor obedience of the
hemispheric sign rule of both magnetic and current helicity.
We argued before that magnetic helicity should be much less affected
by noise than the current helicity, but this is not supported by
the current data.
The reason for this is not obvious.
Looking, for example, at the case of NOAA~10930, we see that the spectrum
is not actually very noisy, but that it has the same sign at almost
all wavenumbers.
Moreover, NOAA~10930 was located in the southern hemisphere, but its
magnetic helicity had the same sign as that normally expected for the
northern hemisphere.
This may then suggest that the hemispheric sign rule violations are
not connected with measurement uncertainties, but they may instead
be physical.
While this is a plausible proposal, it remains curious as to why
much weaker fluctuations are generally seen at poorer resolution.
One possibility is that there are significant systematic errors that tend
to produce magnetic helicity in agreement with the hemispheric sign rule.
Extreme evidence for this comes from the results of the analysis
of synoptic magnetograms \citep{BPS17,Singh18}, where very little
departure from the hemispheric sign rule has been found.
In some cases, those results with poorer resolution showed even
wavenumber-dependent sign reversals of magnetic helicity that agreed
with theoretical expectations \citep{Singh18}.

Such an interpretation in which measurement errors would display a
systematic hemispheric dependence would be difficult to accept.
It would also raise the question of what is the nature of such systematic
errors that produce, or reproduce, the expected sign role.

Another interpretation of our results could be that the measurements at
poorer resolution are actually real, but that the effect of individual
active regions becomes subdominant at poorer resolution.
The high resolution {\em Hinode} images, on the other hand, resolve significant
detail, which makes their contribution dominant.
If this is true, we must accept that the magnetic helicity of individual
active regions can significantly deviate from the hemispheric sign
rule, while the more diffuse background field obeys the potentially
scale-dependent hemispheric sign rule rather well.
This interpretation is further supported by a recent reanalysis of synoptic
vector magnetograms using the parity-even and parity-odd $E$ and $B$ mode
polarizations \citep{Bra19}, which avoids the uncertainty associated with
the $\pi$ ambiguity.
Those results showed a much clearer sign reversal toward low wavenumbers.
This was interpreted to be due to contributions that are far from active
regions, where the $\pi$ ambiguity is more problematic in conventional
methods.

\acknowledgments
{\em Hinode} is a Japanese mission developed and launched by ISAS/JAXA,
collaborating with NAOJ as a domestic partner, and NASA and STFC (UK)
as international partners.
Scientific operation of the {\em Hinode} mission is conducted by the {\em Hinode}
science team organized at ISAS/JAXA.
This team mainly consists of scientists from institutes in the partner
countries.
Support for the post-launch operation is provided by JAXA and NAOJ
(Japan), STFC (U.K.), NASA (U.S.A.), ESA, and NSC (Norway).
We thank Maarit K\"apyl\"a, Alexei Pevtsov, Ilpo Virtanen, and Nobumitsu
Yokoi for providing a splendid atmosphere at the Nordita-supported
program on Solar Helicities in Theory and Observations.
We also thank the anonymous referee for constructive remarks.
This work was supported in part through the National Science Foundation,
grant AAG-1615100, and the University of Colorado through its support
of the George Ellery Hale visiting faculty appointment,


\end{document}